\documentclass{mem}
\usepackage{natbib}
\usepackage{graphicx}
\usepackage[a4paper]{hyperref}
\idline{XXXX}{1}
\begin{document}
\def\teff{$T\rm_{eff }$}
\def\kms{$\mathrm {km s}^{-1}$}
\def\ltsima{$\; \buildrel < \over \sim \;$}
\def\simlt{\lower.5ex\hbox{\ltsima}}
\def\gtsima{$\; \buildrel > \over \sim \;$}
\def\simgt{\lower.5ex\hbox{\gtsima}}

\title{Lessons from Surveys of The Galaxy
}

   \subtitle{}

\author{
Rosemary F.G.~Wyse 
          }

  \offprints{R.~Wyse}

\institute{
The Johns Hopkins University, Department of Physics \& Astronomy, Baltimore, MD 21218, USA 
\email{wyse@pha.jhu.edu}
}

\authorrunning{Wyse}

\titlerunning{Surveys of The Galaxy}

\abstract{
I present a brief survey of surveys, and their results and implications, intended to set the context for the ensuing discussion. 
}
\maketitle{}

\section{Introduction}

These are exciting times to study local galaxies, due to the confluence of three approaches: 

{$\bullet$} 
Advances in technology have allowed  large, high-resolution simulations of structure formation to 
model   Galaxy formation in a cosmological context 

{$\bullet$} Large observational surveys of stars in Local Group galaxies are now possible, using wide-field imagers and multi-object spectroscopy, complemented by space-based imaging and spectroscopy, followed in the near future by  the GAIA satellite and full phase space information 

{$\bullet$} High-redshift surveys are now quantifying the stellar populations and
morphologies of galaxies at high look-back times
 

\noindent I will here focus on the second approach, while
acknowledging the synergy with the others.  I will not attempt a full
historical review, but highlight advances with what I consider
appropriate examples.  I will also focus on {\it stellar\/}
components, but one should not forget the importance of the
interstellar medium. 

\section{Early Surveys}
 
\subsection{Star Counts}
 
Since the work of Kapteyn in the early 20th centuries, star
counts, particularly at high Galactic latitude, have been utilised to
define Galactic structure.  However, their shortcomings, when taken
alone, have also been long known.  The apparent magnitude distribution
of stars depends on many factors, not only their density distribution
-- their luminosity function depends on metallicity, their birth-rate
and the underlying (invariant?) initial mass function (see e.g.~Gilmore \& Wyse 1987 for a review).
Refining star counts to include colour allows more stringent testing
of models, but again the result is critically dependent on the adopted
luminosity function and colour-magnitude relation for different
populations/components in the model (this may seem obvious, but the use of inappropriate choices was the source of much contentious debate in the 1980s). 

Bearing their limitations in mind, star counts in selected
lines-of-sight proved extremely useful for delineating the overall
large-scale structure of the stellar components of the Galaxy. This is
usually achieved not by direct inversion of the star counts, but by
comparisons of the observations with model predictions (van den Bergh
1980; Bahcall \& Soneira 1980; Gilmore 1981).  In particular, the
stellar density laws (radial and vertical) of the thin disk were
derived (Sandage \& Katem 1977; Bahcall \& Soneira 1981; Yoshii 1982); the density profile
and shape of the stellar halo were estimated from a variety of tracers
(e.g. Hartwick 1987; Wyse \& Gilmore 1989; Reid \& Majewski 1993;
Kinman, Suntzeff \& Kraft 1994); the thick disk was defined as a
component with exponential scale height some 3--4 times that of the
old thin disk (Gilmore \& Reid 1983; Fenkart 1988; Larsen \& Humphreys 2003).
 
The interpretation of early star counts was complicated by several
factors. Degeneracies in reddening--age--metallicity were exacerbated
by the fact that the counts were based on photographic photometry, in
only a limited range of bandpasses.  Poor star-galaxy separation at
faint magnitudes ($V \simgt 20$) can cause problems, particularly for blue
objects (see the discussion in Reid \& Majewski 1993).  The small number of lines-of-sight
in any one survey and limited areal coverage further made it difficult
to isolate the underlying cause(s) of discrepancies between different 
investigations. 

Determination of metallicities greatly aids the interpretation. Those
derived from  photometry can of course be obtained for more stars with less
investment of telescope time, compared with spectroscopic determinations.  The broad-band (UBV) based metallicity
distributions of faint F/G-dwarfs by Gilmore \& Wyse (1985), combined with density laws derived from star counts,  were 
critical in ascertaining that the thick disk was indeed a distinct
component.  Intermediate and narrow-band photometry such as Str\"omgren
photometry remains an effective tool, most recently illustrated by the
work of Nordstr\"om et al.~(2004), providing the definitive analysis of
the metallicity distribution of nearby F/G dwarfs.

\subsection{Star Counts plus Kinematics}

Photometric observations repeated after a sufficiently long baseline
allow for the combination of star counts plus proper motions.  The
reduced proper motion diagram is a useful discriminant of different
kinematic populations, in the absence of reliable distances.  Chiu
(1980) applied this technique to his database of proper-motions in
Cardinal directions\footnote{The Galactic poles, the
anti-center/center line and towards/away from Galactic rotation} for a
faint ($V \simlt 20.5$) magnitude-limited sample, based on deep photographic plates with a
25~year baseline.  He  concluded that `Population~I' far from the disk
plane was of lower metallicity, and had higher velocity dispersions,
than did Population~I locally.  A re-analysis of his data, without the
requirement that there be only {\it two\/} components (the classical Populations I and II), showed the
presence of the intermediate-kinematics thick disk (Wyse \& Gilmore
1986).

Distances derived from photometric parallaxes plus proper motions, for
stars selected purely as a magnitude-limited sample, i.e. not kinematically
selected, in at least one of the Cardinal directions, can be used to
probe one or more components of the space motion (e.g.~Majewski
1992), as can radial velocities (Sandage \& Fouts 1987b). Reliable distances based on photometry need good metallicity
and gravity estimates, which are not always available.  The Hipparcos/Tycho sample (with trigonometric parallaxes) allowed an analysis
of very local space motions, such as the velocity dispersion tensor as a function of colour, on the scale of less than a hundred parcsecs, of course
for mostly disk stars (e.g.~Dehnen \& Binney 1998).

The combination of star counts plus spectroscopy -- to provide a metallicity estimate in
addition to radial velocity -- is more powerful than photometry alone,
allowing joint analyses of metallicity and kinematics/dynamics.  For
example, this combination allows a robust dynamical analysis of local
vertical motions to derive the vertical acceleration and associated
mass density and surface density ($K_z$; Kuijken \& Gilmore 1989).
The conclusion from this analysis is that there is no `extra'
dissipative dark matter, confined to the disk, in additional to the
dissipationless dark matter in the (dark) halo.

Radial velocities, plus distances and proper motions, allow full
phase-space investigations. Early surveys used photometric parallaxes
for distances, plus a minimum proper motion selection criterion to
define the samples, in order to increase the `yield' of non-thin disk
stars in these necessarily local sample.  This kinematic selection
requires an understanding of, and correction for, the kinematic bias
introduced.  The analysis is complicated, but fruitful (Sandage \& Fouts
1987a; Carney \& Latham 1987; Carney, Laird, Latham \& Aguilar 1996).
Again the Hipparcos/Tycho sample provided the opportunity for the
derivation of space motions for local stars, without proper-motion
selection, once metallicities and radial velocities were obtained
(Nordstr\"om et al.~2004).
 
The modern era of star counts derived from very wide-field CCD
photometry such as available from the Sloan Digital Sky Survey
(SDSS, e.g. Chen et al.~2001; Ivezi\'c's and Newberg's talks in these
proceedings) was preceded by pencil-beam CCD photometry in selected
areas, providing multi-band, deep, data over several, to many, square
degrees (e.g. Phleps et al.~2000; Siegel et al.~2002).  These pointed
to tantalizing inconsistencies in different fields. 
\section{Motivation for Surveys: Cosmology}

\subsection{Early Surveys}

The idea that the stellar populations of the Milky Way Galaxy have
critical importance for understanding larger issues in cosmology has
been a major motivation for decades.  Much fundamental early work on
stellar populations and Galaxy formation by Sandage (e.g.~Eggen, Lynden-Bell \& Sandage 1962; Sandage 1970) was centered
around the questions of `how old are the oldest stars in the Galaxy,
how long after the Big Bang did galaxies initiate their collapse, and
what was the duration of that collapse?'  These are crucial in 
constraining the age of the Universe -- obviously as least as old as the
oldest stars -- and hence the values of cosmological parameters, particularly when these are estimated through comparison with measurements of the present value of the Hubble constant. 
The fact that the early, rapid collapse model developed in 
Eggen, Lynden-Bell \& Sandage (1962) is still used as a paradigm for the formation of the Milky Way Galaxy is testament to its simplicity and power. 
 
Significant impetus to use the Milky Way as a template for testing
theories of galaxy evolution also came from inconclusive attempts to derive
cosmological parameters from observations of galaxies, such as the
Hubble diagram (apparent magnitude {\it vs\/} redshift), galaxy number
counts etc.  It was then realised that the evolution of galaxies must be
understood first (summarised in Tinsley 1977), and the Milky Way was potentially an ideal testbed.  The drive to
understand the age distributions and metallicity distributions of the
different stellar components of the Galaxy led to elegant analyses of
the metallicity distribution of long-lived stars, manifest in the local G-dwarf metallicity distribution (Pagel \& Patchett 1975,
van den Bergh 1962; Schmidt 1963) and predictions for the chemical evolution of the Galaxy
beyond the local disk.  The simple, closed-box model of chemical
evolution had been developed (e.g.~Schmidt 1963; Searle \& Sargent 1972) 
and the application to the local disk revealed a `G-dwarf problem' in that the model significantly over-predicted the metal-poor tail of the metallicity distribution of long-lived stars. Analytic and numerical models of
chemical evolution  showed the several ways in which the G-dwarf `problem' could be solved (e.g.~Tinsley 1975; Pagel \& Patchett 1975).  These models included such currently topical aspects as inhomogeneities, and the
interpretation of the pattern of elemental abundances, showing how they trace the past stellar
Initial Mass Function and star formation history (Tinsley 1976; 1979). The data did not merit a full exploitation of these insights.

\subsection{The Modern Era}

Significant motivation for study of Galactic populations still comes
from cosmology, but now not so much the estimation of the present
values of cosmological parameters such as the deceleration parameter
$q_0$ or Hubble constant $H_0$, but the testing of predictions of
galaxy formation in the context of particular cosmological models. The favoured
model at present is $\Lambda$CDM ($\Omega_\Lambda \simeq 0.7$, $\Omega_{matter} \simeq 0.3$, $H_0 \simeq 70$~km/s/Mpc), based on the excellent agreement of its predictions with
measurements of large-scale structure,  such as the fluctuations in the
cosmic microwave background (Spergel et al.~2006) and the galaxy power
spectrum (Sanchez et al.~2006; Eisenstein et al.~2005).  As is
well-known, such a model predicts that large galaxies such as the
Milky Way form and evolve through the merging and accretion of smaller
systems, with the `first objects' having a mass of perhaps $\sim
10^6~$M$_\odot$  (the characteristic mass, and the relation of
these objects to present-day dwarf galaxies is the subject of much
on-going work).  As is also well-known, this model faces several
challenges, particularly concerning its predictions on the scales of groups of
galaxies and below.

The merging history of a typical massive-galaxy dark halo is fairly
straightforward to calculate, since only gravity is involved. However,
most simulations lack the resolution to follow how far inside a
`parent' halo a merging satellite penetrates, and this is crucial to
determine the effect on the baryonic disk.  During mergers the orbital
energy goes into the internal degrees of freedom of the merging
systems, thus `heating' them.  A corollary is that surviving dark
substructure, as predicted in CDM simulations (Moore et al.~1999;
Klypin et al.~1999), can also heat thin disks.  Thin disks are thus
fattened, and while gas can cool and re-settle to the disk plane,
stellar disks remain `hot'.  During a `minor' merger (mass ratio of
less than $\sim 1:4$), the (relatively) low density, outer regions of
the smaller system are removed by tides, to be absorbed into the
larger system.  Orbital angular momentum is also absorbed and
redistributed, with in general outer parts gaining angular momentun
and inner parts losing.  In the process gas and stars are driven to
the center, perhaps helped by a bar that is often predicted to form as
a result of instabilities in the disk.  The disk formed subsequently
has a short scale-length: the corollary is that detailed angular
momentum conservation is required in order to form extended disks as
observed (Fall \& Efstathiou 1980).  Various schemes have been
developed to suppress angular momentum transport and redistribution,
usually invoking some `feedback' process to maintain the baryons in a
diffuse gaseous state for as long as possible (e.g.~Weil, Eke \& Efstathiou 1998; Maller \& Dekel
2002; Robertson et al.~2006), until the epoch of active (major) merging is complete, perhaps even as recently as a redshift of unity. 

Abadi et al.~(2003) present a recent simulation
of the formation of a present-day disk galaxy that demonstrates many
of the important aspects, including the outstanding problem of how to
include star formation and gas physics.  Generic predictions for disk
galaxies include the following:

\begin{itemize}

\item{} Extended disks settle and form late, after the last major mergers, typically (for a dark halo of mass $10^{12}$M$_\odot$) corresponding to a redshift of unity (e.g. Maller et al.~2006), or a
lookback time of $\sim 8$~Gyr

\item{} A large disk galaxy should have hundreds of surviving
satellite dark haloes at the present day; these may well provide
observable signatures through their gravitational interactions with
the baryonic galaxy, such as heating of the thin disk, disruption of
wide binaries, disturbance of extended tidal streams etc.

\item{} The stellar halo is formed from disrupted satellite galaxies  

\item{} Minor mergers (a mass ratio of $\sim 20$\% between the
satellite and the disk) into a disk continue after the last `major merger', and heat it, forming a thick disk out
of a pre-existing thin disk, and create torques that drive gas into
the central (bulge?) regions

\item{} More significant mergers transform a disk galaxy into an S0 or
even an elliptical

\item{} Subsequent accretion of gas can reform a thin disk 

\item{} Stars can be accreted into the thin disk from suitably massive
satellites (dynamical friction must be efficient) and if to masquerade
as stars formed in the thin disk, must be on suitable high angular
momentum, prograde orbits

\end{itemize}

\subsection{The Fossil Record: Tests of Predictions}

Stars of mass like the Sun, and lower, live for essentially the age of
the Universe, and retain memory of many aspects of the conditions at
early times.  Studying old stars nearby thus allows us to study 
cosmology locally, a very complementary approach to direct study of
high-redshift objects. There are copious numbers of stars in Local
Group galaxies that have ages of greater than 10~Gyr, and thus formed
at lookback times equivalent to redshifts of 1.5 and greater (see Figure~1).

The clues to galaxy evolution that one might wish to extract from the
local fossil record include the star formation history, the form of
the stellar Initial Mass Function and whether or not it varied between
then and now, chemical evolution, and the relative importances of
dissipative gas physics versus dissipationless processes. The overall dark halo potential well depth and shape can be 
inferred from stellar (and gas) kinematics.   There are
aspects of the stellar populations that are less sensitive to details of 
baryonic
physics -- such as the ages of stars in the thick disk -- and these can be used to constrain the merging history -- is
this compatible with $\Lambda$CDM? 
Is the Milky
Way a typical galaxy?  

Most galaxies in the local Universe are observed to cluster in loose
groups like the Local Group (which in itself is unusual in CDM models, 
Governato et al.~1997). While lacking a giant elliptical, the
Local Group hosts a reasonably diverse selection, with large disk
galaxies of a range of bulge-to-disk ratios (The Milky Way, M31, M33),
gas-rich and gas-poor satellites ranging from the compact elliptical
M32 through the numerous extremely low surface brightness dwarf
spheroidals (dSph). Do trends in inferred merging history etc for 
the Local Group galaxies match predictions? 

We can address these questions with current and planned capabilities, with which the motions, spatial
distributions, ages and chemical elemental compositions can be
measured (with varying accuracies!) for individual stars in galaxies
throughout the Local Group, plus additional complementary tracers such as
HII regions and planetary nebulae. 
 
What have we learnt so far?

\section{Milky Way Large Scale Structure}
 
 \subsection{The Thin Disk}

The large-scale structure of the thin stellar disk is reasonably well modelled
by a double exponential with scalelength of $\sim~3$~kpc and
scaleheight of $\sim~300$~pc (for stars older than a few Gyr).
Extrapolating this smooth structure with a local normalization  for 
stellar surface density of $\Sigma_*~\sim~35$M$_\odot$~pc$^{-2}$ (Kuijken \& Gilmore 1989; Flynn et al.~2006) gives a
total mass of around $6 \times 10^{10}$M$_\odot$.  The interstellar
medium contributes $\sim 10$M$_\odot$~pc$^{-2}$ locally, and has a rather different radial profile from the stars, with atomic and molecular gas each having a distinct spatial distribution.
 
Stellar metallicity and age distributions are best-known at present for the local
disk, within around one kpc of the solar circle.  As noted above,
Str\"omgren photometry has proven a robust technique of metallicity
determination for large samples of F/G dwarfs, confirming the `G-dwarf
problem' in the local disk i.e.~a narrow metallicity distribution,
with few stars significantly more metal-poor than the peak, in
contradiction to the large metal-poor tail predicted by the simplest
chemical evolution models (e.g.~Wyse \& Gilmore 1995; Rocha-Pinto \&
Maciel 1996; Nordstr\"om et al.~2004).  The peak metallicity of
long-lived stars in the solar neighbourhood is somewhat below the
solar value, $\sim -0.15$~dex, with good agreement between 
G-dwarfs and lower-mass K-dwarfs (Kotoneva et al.~2002). High-resolution spectroscopic studies of necessarily smaller samples provides a peak iron abundance of $\sim -0.1$~dex (Allende-Prieto et al.~2004).  

The star formation history of the {\it local\/} stellar  disk has been estimated
through various techniques, and the general conclusion is for an early
onset, an approximately constant overall rate, and
with low-amplitude (factor of two) bursts on (few?) Gyr timescales
(e.g.~Hernandez, Valls-Gabaud \& Gilmore 2000; Rocha-Pinto et
al.~2000). There is certainly no lack of old stars in the local thin
disk, a location that is some 3 scalelengths from the Galactic center.
Assuming these stars formed in the thin disk, one concludes that an
extended thin disk was in place at a redshift of around 2
(corresponding to the look-back time of 10--12~Gyr estimated for the
onset of star formation in the local disk; Binney, Dehnen \& Bertelli
2000). This is significantly earlier than a typical extended thin disk would form in
CDM models.  

Complementary data for external disk galaxies of similar
scale-length to the Milky Way (half-light radii of between 5kpc and 7kpc) show little
evolution in size or number back to a redshift of unity (the limit of
the data, for the COSMOS survey sample of Sargent et
al.~2006). Hence, extended disks do not seem to {\it start\/} forming at $z
\sim 1$, but rather to be well-established by then.  Models in which a
significant part of the old thin stellar disk is formed by the later addition of old stars by satellite
accretion directly into high-angular momentum orbits in the disk plane
(e.g.~Abadi et al.~2003) need to address this. 

\begin{figure}[]
\resizebox{\hsize}{!}{\includegraphics[angle=270]{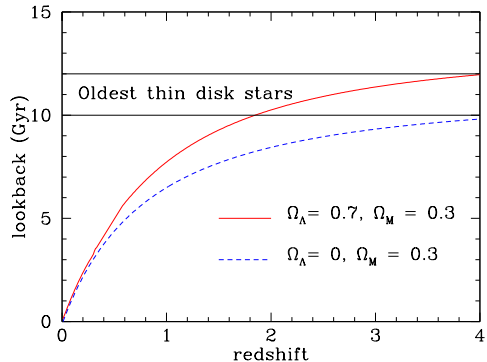}}
\caption{ \footnotesize Plot of lookback time versus redshift for two cosmological models, as labelled, each with present value of the Hubble constant of 70~km/s/Mpc.  The horizontal lines indicate estimates of the ages of the oldest stars in the thin disk, from Binney et al.~(2000).}
\label{thinage}
\end{figure}

 \subsection{The Thick Disk}

The large-scale structure of the thick stellar disk is (probably!) reasonably well modelled
by a double exponential, with scalelength of $\sim 3$~kpc and
scaleheight of $\sim 1$~kpc, giving an axial ratio that is  a factor of three or so `fatter' than the thin disk.  
Extrapolating this smooth structure with a local normalization of around 5\% gives a total mass of around 15\% of that of the thin stellar disk. 

Again the metallicity and age distributions are best determined at
present only fairly locally, within a few kpc of the Sun, both
vertically and radially.  The tails of the derived kinematic and
metallicity distributions overlap with those of the thin disk, so
there is the danger in very local samples of being overwhelmed by the
much more numerous thin disk stars.  Defining a thick disk sample {\it
in situ\/}, for example above $\sim 1$~kpc vertically from the disk
plane, provides an effective filter. Such samples find a thick disk
peak metallicity of $\sim -0.6$~dex, and that essentially all the
stars have an age as old as the globular cluster of the same
metallicity as the stars, or some 10-12~Gyr (e.g.~Gilmore \& Wyse
1985; Gilmore, Wyse \& Jones 1995; see also Ratnatunga \& Freeman
1989; Morrison, Flynn \& Freeman 1990). Local proper-motion selected
samples find similar results (e.g.~Carney, Latham \& Laird 1989),
albeit with alternative interpretations (e.g.~Norris \& Ryan 1991).

The extension of the metallicity distribution of the thick disk to
metallicities significantly below $\sim -1$~dex (e.g. Morrison, Flynn \& Freeman 1990) remains topical,
with its most robust detection in samples selected by having low
metallicity, and thus with uncertain normalization to the main peak
(e.g.~Chiba \& Beers 2000). 

As discussed in Sofia Felzing's contribution, the stars in the local
thick disk follow a distinct elemental abundance pattern, offset from
stars in the thin disk (defined kinematically).  This presumably
reflects the different star formation histories of these two
components, the thick disk having a (significantly) shorter duration
of star formation.  A well-defined separation of populations on the
basis of elemental abundance patterns holds much promise for
identification of substructure and tracing the history of the Galaxy (cf.~Freeman \& Bland-Hawthorn 2002). 
 
The thick disk has kinematics that are intermediate between those of
the thin disk and stellar halo: a typical local thick disk star is on 
a fairly high angular momentum orbit, with a lag behind the mean
azimuthal (rotational) velocity of the old thin disk of only $\sim 30
- 50$~km/s. The vertical velocity dispersion is around 45~km/s, hotter
than can be achieved through heating the thin disk by local
gravitational perturbations such as Giant Molecular Cloud complexes
and/or transient spiral arms. 
 
The dominant old age of stars in the thick disk, $\sim 11$~Gyr,
combined with the large age range of stars in the thin disk, argues
against models in which the thick disk forms from the thin disk by a
heating process that occurs over an extended period.  If the heating
is merger-induced (the minor-merger scenario for formation of the thick disk from a pre-existing thin disk), then the last significant merger into the thin disk
was long ago, at a redshift $\sim 2$, corresponding to a lookback time
of $\sim 11$~Gyr.  This is unusually long ago in $\Lambda$CDM models,
particularly when one remembers that for sufficient heating of the thin disk, the
`significant' merger need only have mass equal to 20\% of the {\it
disk} mass, not the total mass. 
 
In any merger-model for the formation of the thick disk, there will be
a contribution to `the thick disk' from stars removed from the culprit
satellite(s). Indeed, in some models tidal debris from shredded
satellite galaxies is a very significant part of `the thick disk'
(e.g.~Abadi et al.~2003). However, the high peak metallicity of the
thick disk stars suggests that these stars formed within a fairly deep
potential well, particularly given their old age; as an example, while
the LMC has managed to self-enrich to a similar metallicity, [Fe/H]$
\sim -0.6$~dex, this is for stars only a few Gyr old. The putative
satellites in which the majority of thick disk stars formed would have to be
extremely different from those surviving satellites. 
 
The evidence from observations of high-redshift systems is limited,
but there has been a recent detection of what appears to be a
kinematically hot (i.e. expected to be thick) stellar disk forming in a burst of star formation at
a redshift of greater than 2 (Genzel et al.~2006).  More nearby galaxies too appear to contain old thick disks (Mould 2005; Yoachim \& Dalcanton 2005), not dissimilar to that of the Milky Way.  

Identifying the analogue (if one exists) of the Milky Way thick disk in M31 is
complex, due in part to the pervasive inhomogeneities in stellar surface densities (Ferguson et al.~2002) and disparate lines-of-sight with
spectroscopic and deep photometric information.  Is the `spheroid'
component with [Fe/H]$\sim -0.6$ the thick disk in M31 (e.g.~Wyse \& Gilmore
1988)? Or is it more associated with the outer disk (Brown et al.~2006), and contains stars of a wide range of ages, thereby compatible with a more extended merger history than the Milky Way? 

The lower-mass spiral galaxy M33 appears to have had a very quiescent
life, with little evidence for significant mergers or interactions.
While a trend between merging history and total mass is expected in
$\Lambda$CDM, such that lower mass dark haloes have fewer recent
mergers (e.g.~Maller et al.~2006), it remains to be seen if the Milky
Way, M31 and M33 can be produced easily.

\subsection{The Central Bulge}

The smooth structure of the central bulge is mildly triaxial,
i.e.~barred, with axial ratios of $\sim 1:0.35:0.3$ (Bissantz \&
Gerhard 2002). The profile is reasonably well-fit by an exponential,
with scaleheight $\sim 300$~pc, and thus the Milky Way bulge is not a
classical `$r^{1/4}$-bulge' but rather perhaps a `pseudo-bulge', often
found in later-type spiral galaxies (e.g.~Carollo, Stiavelli \& Mack
1998).  The total stellar mass of the bulge is $\sim
10^{10}$~M$_\odot$, and the central regions are very baryon-dominated
(Bissantz, Debattista \& Gerhard~2004).
 
The stellar populations have been studied mostly in `windows' of low
optical extinction.  The peak spectroscopic metallicity from samples of K-giants is
somewhat below the solar value (e.g. McWilliam \& Rich 1994; Ibata \& Gilmore 1995; Fulbright, McWilliam \& Rich 2006a), similar
to the long-lived stars in the local thin disk.  The dominant age is
old, again 10-12~Gyr, with younger stars in lower latitude central
regions (e.g.~Ortolani et al.~1995; Feltzing \& Gilmore 2000; Kuijken \& Rich 2002; van Loon et al.~2003).  The coincidence in age with the thick
disk may point to one merger event to set the physical conditions for
both components: the gas driven to the centre during the merger that
heated the thin disk to form the thick disk, would form the bulge
(e.g.~Wyse 2001).
 
Determinations of elemental abundances are limited to small samples of
K-giants (typically around 50 stars), and are consistent with enrichment by (normal IMF) Type II
supernovae only, i.e.~the stars formed in only a short duration of star
formation (e.g.~Fulbright, McWilliam \& Rich 2006b; Zoccali et
al.~2006).  This short duration agrees with earlier inferences from more limited data (e.g.~Matteucci \& Brocato 1990; Rich 1999; Ferreras, Wyse \& Silk 2003).

The low-mass end of the IMF in the bulge can be studied by direct star
counts, with the result (Zoccali et al.~2000) that it is
indistinguishable from that in (dynamically unevolved) metal-poor
globulars.  The low-mass IMF in the Ursa Minor dwarf Spheroidal galaxy
(Wyse et al.~2002) is also indistinguishable from that in metal-poor
globular clusters, and again the elemental abundances in dSph do not
require any variations in massive-star IMF. The IMF seems remarkably
invariant with metallicity, epoch of star formation, (present) stellar
density etc.

Models of how the bulge in the Milky Way formed generally appeal either to its
`pseudo-bulge' density profile and triaxial shape to argue for an
instability in the inner disk (in which case the formation of the
bulge could have occured significantly after the formation of the
stars themselves), or to its rapid enrichment and high density 
to argue for an {\it in
situ} starburst (e.g.~Elmegreen 1999; see the review in  Wyse 1999).

\subsection{The Stellar Halo}

The large-scale structure of the inner regions, within $\sim 15$~kpc of the
Galactic center is the best-constrained at present. The dominant
population is old and metal-poor, with the stars on low angular
momentum orbits. The overall density profile (traced by RR~Lyrae
stars) shows a smooth power-law fall-off with distance (measured in
the disk plane) of $\rho_{RRL} \propto R^{-3.1}$ out to $\sim 50$~kpc
(Vivas \& Zinn 2006). The stellar halo as traced by main sequence F/G
stars and RR Lyrae stars is not spherical, but can be reasonably well fit by
an oblate spheroid, with flattening at around the solar distance of
$c/a \sim 0.5$ (Hartwick 1987; Wyse \& Gilmore 1989) becoming rounder
with distance, and approximately spherical at $R \simgt 20$~kpc (Vivas
\& Zinn 2006).  The total stellar mass in this smooth distribution is
$\sim 2 \times 10^9$~M$_\odot$ (e.g.~Carney, Latham \& Laird 1990).
Hints of triaxiality are seen in deep imaging data, as discussed in the meeting by Heidi Newberg.

Both the age distributions (Unavane, Wyse \& Gilmore 1996) and the
elemental abundance patterns (Fulbright 2002; Stephens \& Boesgaard 2002; Tolstoy et al.~2003; Venn et al.~2004) of the
bulk of the field halo stars are very different from those in the
present satellite galaxies of the Milky Way (with the abundance
pattern consistent with the expectations from the extended star
formation histories).  Accretion of stars from systems like the
satellite galaxies, into the field halo, is limited to less than 10\%  since a redshift of unity (Unavane et al.~1996).  The rare
halo stars with extremely high velocities, probing the outer halo,
have lower values of [$\alpha$/Fe], more similar to the stars in the
dSph, but the overall abundance pattern remain different (Fulbright
2002).

The abundance ratios of lighter metals in the field halo stars show remarkably
little scatter down to the lowest metallicities (e.g. Cayrel et
al.~2004), defining a flat `Type II plateau' in [$\alpha$/Fe] and indicating that the stars
formed in systems with only a short duration of star formation, 
allowing enrichment by only the short-lived progenitors of Type ~II
supernovae.  The value of a predicted `Type~II plateau' in [$\alpha$/Fe] depends on the massive-star IMF (see e.g. Wyse \& Gilmore 1992), and the low amplitude of scatter indicates an invariant IMF. 

The mean metallicity of the stellar halo is around $-1.5$~dex (e.g.~Ryan
\& Norris 1991), significantly lower than the local (gas-rich) disk. 
With a fixed
stellar initial mass function, and no gas flows, one expects a system
of low gas fraction, such as the
stellar halo,  to be {\it more\/}
chemically evolved than a system with higher gas fraction.
Hartwick (1976) provided an elegant explanation to this conundrum: gas outflows from active star-forming regions in the proto-halo. 
The chemical evolution requirements are such that for
a fixed stellar IMF, one that matches the local thin disk mean metallicity of just below the solar value, the outflows must occur at around 10 times the
rate of star formation.  An attractive corollary to this picture is
that one can tie the gas outflow from low-mass halo star-forming regions to gas
inflow to the central regions to form the bulge; the low angular
momentum of halo material means that it will only come into
centrifugal equilibrium after collapsing in radius by a significant
factor.  The mass ratio of bulge to halo is around a factor of ten,
just as would be expected, and the specific angular momentum
distributions of stellar halo and bulge match (Wyse \& Gilmore 1992;
see Figure~2 here).
 
\begin{figure}[]
\resizebox{\hsize}{!}{\includegraphics[angle=0]{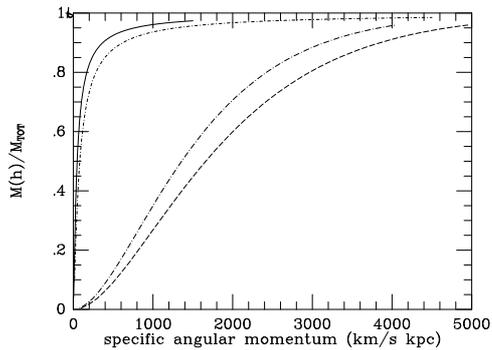}}
\caption{ \footnotesize Adapted from Wyse \& Gilmore 1992, their Figure~1.  Angular
momentum distributions of the bulge (solid curve), the stellar halo
(short-dashed/dotted curve), the thick disk (long-dashed/dotted curve)
and the thin disk (long-dashed curve).  The bulge and stellar halo
have similar distributions, as do the thick and thin disks.}
\label{angmom}
\end{figure}

\section{Small-Scale Structure}

\subsection{The Outer Stellar Halo}

The outer halo, with dynamical timescales of $> 1$~Gyr, is the best
location to search for structure. Indeed, several streams have been
found, in both coordinate space and in kinematics. Most of these
appear to be due to the Sagittarius Dwarf Spheroidal (Sgr dSph), a
galaxy that was discovered serendipitously during a survey of the
kinematics and metallicity distributions of stars in the Milky Way
bulge (Ibata, Gilmore \& Irwin 1994, 1995).  Extended tidal streams
from the Sgr dSph are now detected across the sky (Ibata et al.~2001;
Majewski et al.~2003; Belokurov et al.~2006a).  These are potentially
extremely useful in constraining the shape and smoothness of the dark
halo potential, although we are in the interesting situation of
contradictory conclusions from different datasets (e.g.~Helmi 2004;
Johnston, Law \& Majewski, 2005;  Fellhauer et al.~2006). 
 
The detection and characterization of structure in the stellar halo is
a very fast-moving field!  As described further in Heidi Newberg's talk in this volume, several (of order 10) candidate new dSph,
globular clusters and streams (including at least one stream from a disrupting globular
cluster --  not all streams indicate accretion from an external source)
have been announced this year, all exploiting the Sloan Digital Sky
Survey imaging data (e.g.~Belokurov et al.~2006a,b; Grillmair 2006).  Determining the {\it masses\/} of the new
putative satellite galaxies is crucial before one can say what impact
these discoveries have on the `satellite problem' of $\Lambda$CDM
models -- namely the over-prediction by the models of satellite-galaxy mass dark
haloes.  At present, the available radial velocity data internal to
the dSph companions of the Milky Way are consistent with each dSph
being embedded in a dark halo of {\it fixed mass\/}, of $\sim 4 \times
10^7$~M$_\odot$ (Wilkinson et al.~2006), out to the extent of the
stars (saying nothing about the extent, or mass, at radii beyond
available stellar kinematic data).  While it is reasonably easy,
theoretically, to modify the baryon content of shallow potential well
systems such as the satellite haloes and thus change their luminosity function, it is much harder to 
conceive of modifications to the predicted mass function of satellite
haloes to match a narrow range  in mass.

\subsection{The Thick Disk}

As noted above, a popular model for the formation of the thick disk is
based on a minor merger into a pre-existing thin disk, with the
orbital energy in large part going into heating the thin disk.  The
satellite galaxy (or galaxies) responsible for the heating does not
survive unscathed, but will in general be tidally disrupted, the mass
of any surviving remnant being set by how deeply it penetrates and the relative density compared that of to
the larger (Milky Way) galaxy.  The `shredded satellite' stars will
likely contribute to the stellar populations above the thin disk
plane, and thus be included in samples of `thick disk' stars.  The
spatial distribution of satellite debris will reflect the orbit of the
satellite, and its (relative) density profile (e.g.~Huang \& Carlberg
1997, their Fig.~19; Abadi et al.~2003).

Such satellite debris has been identified, on the basis of distinct
kinematics, namely lower angular momentum than the bulk of thick disk
stars, in at least one sample of turn-off stars observed several kpc
from the disk plane (Gilmore, Wyse \& Norris 2002).  

The details of the stellar populations in the thin disk -- thick disk
-- halo transition region contains much information about the past
merger history of the Milky Way. Large samples (many thousands) of
stars are needed; we (Gilmore, Wyse \& Norris) have undertaken a
moderate survey with the new multi-object spectrograph on the
Anglo-Australian Telescope, AAOmega, targeting $\sim 13,000$ F/G stars in the
equatorial stripe of the imaging data of SDSS DR4.  Results from the SDSS spectroscopic database are presented by Ivezi\'c in this volume.  The Galactic structure survey of the
Sloan Digital Sky Survey Extension (SDSS-II/SEGUE) is also on-going.

 \subsection{The Thin Disk}
 
The possibility that a significant fraction of
old stars found now in the thin disk were formed in satellite galaxies that were subsequently accreted into the disk plane was suggested by the simulations of  Abadi et al.~(2003).  A typical
satellite orbit is far from circular in the disk plane (e.g.~Benson 2005) so that this
scenario requires that the satellites be massive enough for dynamical
friction to damp the vertical orbital motion and circularize their orbit quickly
enough, and it remains to be seen if, for example, the chemical composition of the satellite stars matches those observed in the old disk. 

Ongoing large-scale spectroscopic surveys such as RAVE (targeting
bright stars with the UK~Schmidt Telescope of the Anglo-Australian Observatory and 6dF multi-object
spectrograph; Steinmetz et al.~2006) 
and SDSS-II/SEGUE should provide ideal databases for an
identification of kinematic substructure in the disk. The large
Geneva/Copenhagen survey of the local disk (Nordstr\"om et al.~2004) is
also a fertile hunting ground, with possible ancient substructure
identified by Helmi et al.~(2006).  The high-resolution mode of the
proposed multi-object spectrograph (WFMOS) for Gemini will provide
unprecedented elemental abundance data, containing much more
information than overall metallicity, and with signatures that persist
longer than spatial or even most kinematic features.

The `ring' around the Galaxy seen in star counts (e.g.~Yanny et
al.~2003; Ibata et al.~2003) could be either a remnant of satellite
accretion into the plane of the disk (e.g.~Bellazzini et al.~2006 and references therein) 
or  more simply structure
in the outer stellar disk, which most probably warps and flares
(e.g.~Momany et al.~2006 and references therein).  Indeed the rich structure in HI gas in the
outer disk may be seen in the Leiden/Argentine/Bonn HI survey
(Kalberla et al.~2005). Even the old disk is unlikely to be well-fit
by a smooth model, given the strong spiral structure seen in K-band
images of external spirals (Rix \& Zaritsky 1995).  

Indeed, the fact that 
the underlying potential of the
disk is neither smooth, axisymmetric nor time-independent cannot be ignored.  
As
demonstrated by De Simone, Wu \& Tremaine (2004)\footnote{Their analysis was 
based on a shearing sheet, only the $z=0$ plane.  A 3D analysis would be very interesting.}, transient
perturbations, such as segments of spiral arms, not only heat the
stellar disk, but can produce `moving groups' that persist long after the
gravitational perturbation has gone.  These kinematic features are
created from random collections of disk stars, and so will contain a
range of ages and metallicities.  Interestingly, such moving groups have been identified (Famaey et al.~2004).  

\section{Survey Requirements}

I hope my brief survey of surveys has demonstrated that the field of
`Galactic Structure' is vibrant and exciting. While much has been
learnt, much remains to be learnt. Future large surveys are needed to
quantify both small-scale and large-scale structures.  These surveys
should be based on input catalogues with excellent and uniform
multi-band  photometry and also excellent astrometry, to give
accurate and precise positions and proper motions.  One can envisage a
hierarchy of surveys, each providing input to the next level.
 
Immediately from imaging data one can analyse the spatial structure, 
in colour/magnitude space. Spectroscopic targets can also be defined, with a wel-understood and uniform selection function across the sky.  
 
Medium resolution spectroscopy (few \AA\  resolution), providing radial
velocities to a few km/s and metallicities to $\sim 0.2$~dex, should
be obtained for several hundreds of stars in each each line-of-sight,
allowing the analysis to go beyond the means and dispersions of
present surveys, and to look at the structure in the kinematic and
metallicity distributions. Metallicity estimates are necessary for improved photometric parallaxes and space motions from proper motions. The sampling strategy -- for example,
sparse sampling, or overlapping fields?  Stripes in longitude or one
large contiguous area at high latitude, etc -- plus the selection
function -- K-giants?  F/G dwarfs? -- should be tuned to the science
goals.
 Observations in the IR may be required to probe the bulge,  plus the lowest-latitude disk.  
 
High resolution spectra should be obtained for the brighter stars, for
elemental abundances and precise velocities. This allows mapping of
substructures defined by kinematics and star formation
history/chemical evolution.  Again, large samples are required, and a
multi-object spectrograph such as the proposed WFMOS Gemini instrument
is ideal.  This is discussed further in this volume by Joss
Bland-Hawthorn.

These are indeed exciting times to study stars in Local Group galaxies.

\begin{acknowledgements}
I am grateful to Chris Corbally and colleagues on the 
Organising Committee for conceiving this stimulating Joint Discussion. 
\end{acknowledgements}

\bibliographystyle{aa}

\begin{thebibliography}{}
\bibitem[]{} Abadi, M., Navarro, J., Steinmetz, M. \& Eke, V.\ 2003, ApJ,  
597, 21
\bibitem[]{} Allende-Prieto, C., Barklem, P.S., Lambert, D. \& Cunha, K.\ 2004, A\&A, 420, 183
\bibitem[]{} Bahcall, J.N. \& Soneira, R.\ 1980, \apjs, 44, 73
\bibitem[]{} Bahcall, J.N. \& Soneira, R.\ 1981, \apjs, 47, 337
\bibitem[]{} Bellazzini, M., Ibata, R., Martin, L., Lewis, G., Conn, B. \& Irwin,
M.\ 2006, MNRAS, 366, 865 

\bibitem[]{} Belokurov, V. et al.\ 2006a, \apj, 642, L137
\bibitem[]{} Belokurov, V. et al.\ 2006b, astro-ph/0608448
\bibitem[]{} Benson, A.\ 2005, MNRAS, 358, 551
\bibitem[]{} Binney, J., Dehnen, W. \& Bertelli, G. 2000, 
     MNRAS, 318, 658
\bibitem[]{} Bissantz, N. \& Gerhard, O.\ 2002, \mnras, 330, 591
\bibitem[]{} Bissantz, N., Debattista, V. \& Gerhard, O. 2004, \apj, 601, L155
\bibitem[]{} Brown, T.M. et al.\ 2006, \apj, in press (astro-ph/0607637)
\bibitem[]{} Carollo, C.M., Stiavelli, M. \& Mack, J.\ 1998, \aj, 115, 2306
\bibitem[]{} Carney, B. \& Latham, D.\ 1987, \aj, 93, 116
\bibitem[]{} Carney, B., Latham, D. \& Laird, J.\ 1989, \aj, 97, 423
\bibitem[]{} Carney, B., Latham, D. \& Laird, J.\  1990, AJ, 99, 572
\bibitem[]{} Carney, B., Laird, J., Latham, D. \& Aguilar, L.\ 1996, \aj, 112, 668
\bibitem[]{} Cayrel, R. et al.\ 2004, A\&A, 416, 1117
\bibitem[]{} Chen, B. et al. (SDSS collaboration) 2001, \apj, 553, 184
\bibitem[]{} Chiba, M. \& Beers, T.\ 2000, \aj, 119, 2843
\bibitem[]{} Chiu, L.-T.~G.\ 1980, \apjs, 44, 31
\bibitem[]{} De Simone, R., Wu, X. \& Tremaine, S.\ 2004, MNRAS, 350, 627
\bibitem[]{} Dehnen, W. \& Binney, J.\ 1998, \mnras, 298, 387
\bibitem[]{} Eggen, O.J., Lynden-Bell, D. \& Sandage, A.R.\ 1962, \apj, 136, 748
\bibitem[]{} Eisenstein, D. et al.\ (SDSS team) 2005, \apj, 633, 560
  \bibitem[]{} Elmegreen, B. 1999, ApJ, 517, 103

\bibitem[]{} Fall, S.M. \& Efstathiou, G.\ 1980, \mnras, 193, 189
\bibitem[]{} Famaey, B. et al.\ 2005, A\&A, 430, 165
\bibitem[]{} Fellhauer, M. et al.\ 2006, \apj, in press (astro-ph/0605026)
\bibitem[]{} Feltzing, S. \& Gilmore, G.\  2000, A\&A, 355, 949
\bibitem[]{} Fenkart, R.\ 1988, A\&AS, 76, 469
\bibitem[]{} Ferguson, A.M.N. et al.\ 2002, \aj, 124, 145
\bibitem[]{} Ferreras, I., Wyse, R.F.G. \& Silk, J. 2003, MNRAS, 345, 1381

\bibitem[]{} Flynn, C., Holmberg, J., Portinari, L., Fuchs, B. \& Jareiss, H.  \ 2006, \mnras, 372, 1149
\bibitem[]{} Freeman, K. \& Bland-Hawthorn, J.\ 2002, \araa, 40, 487
\bibitem[]{} Fulbright, J.\ 2002, \aj, 123, 404
\bibitem[]{} Fulbright, J., McWilliam, A. \& Rich, R.M.\ 2006a, \apj, 636, 821
\bibitem[]{} Fulbright, J., McWilliam, A. \& Rich, R.M.\ 2006b, \apj, in press (astro-ph/0609087)
\bibitem[]{} Genzel, R.\ 2006, Nature, 442, 786
\bibitem[]{} Gilmore, G.\ 1981, \mnras, 195, 183
\bibitem[]{} Gilmore, G. \& Reid, I.N.\ 1983, MNRAS, 202, 1025
\bibitem[]{} Gilmore, G. \& Wyse, R.F.G.\ 1985, \aj, 90, 2015
\bibitem[]{} Gilmore, G. \& {Wyse, R.F.G.}\  1987, in  `The Galaxy', eds G. Gilmore \& R. Carswell. (Reidel : Dordrecht) p247
\bibitem[]{} Gilmore, G., Wyse, R.F.G. \& Jones, J.B.\ 1995, \aj, 109, 1095
\bibitem[]{} Gilmore, G., Wyse, R.F.G. \& Norris, J.E.\ 2002, ApJL, 574, L39 
\bibitem[]{} Governato, F. et al.\ 1997, NewA, 2, 91
\bibitem[]{} Grillmair, C.\ 2006, ApJ, 645, L37
\bibitem[]{} Hartwick, F.D.A.\ 1976, ApJ, 209, 418
\bibitem[]{} Hartwick, F.D.A.\  1987, in  `The Galaxy', eds G. Gilmore
\& R. Carswell. (Reidel : Dordrecht) p281
\bibitem[]{} Helmi, A.\ 2004, \apj, 610, L97
\bibitem[]{} Helmi, A., Navarro, J., Nordstr\"om, B., Holmberg, J., Abadi, M. \& Steinmetz, M.\ 2006, MNRAS, 365, 1309
\bibitem[]{} Hernandez, X., Valls-Gabaud, D. \& Gilmore, G.\ 2000, \mnras, 316, 605
\bibitem[]{} Huang, S. \& Carlberg, R.\ 1997, \apj, 480, 503
\bibitem[]{} Ibata, R. \& Gilmore, G.\ 1995, MNRAS, 275, 605
\bibitem[]{} Ibata, R., Gilmore, G. \& Irwin, M.\ 1994, Nature, 370, 194
\bibitem[]{} Ibata, R., Gilmore, G. \& Irwin, M.\ 1995, \mnras, 277, 781
\bibitem[]{} Ibata, R., Lewis, G., Irwin, M., Totten, E. \& Quinn, T.\ 2001, \apj, 551, 294
\bibitem[]{} Ibata, R.,  Irwin, M., Lewis, G., Ferguson, A. \& Tanvir, N.\ 2003, MNRAS, 340, L21
\bibitem[]{} Johnston, K.V., Law, D. \& Majewski, S.\ 2005, \apj, 619, 800
\bibitem[]{} Kalberla, P.M.W. et al.\ 2005, A\&A, 440, 775
\bibitem[]{} Kinman, T.D., Suntzeff, N.B. \& Kraft, R.P.\ 1994, \aj, 108, 1722
\bibitem[]{} Klypin, A., Kravtsov, A., Valenzuela, O. \& Prada, F.\ 1999, \apj, 522, 82
\bibitem[]{} Kotoneva, E., Flynn, C., Chiappini, C. \& Matteucci, F.\ 2002, \mnras, 336, 879
\bibitem[]{} Kuijken, K. \& Gilmore, G.\ 1989, \mnras, 239, 605
\bibitem[]{} Kuijken, K. \& Rich, R.M.\ 2002, \aj, 124, 2054
\bibitem[]{} Larsen, J. \& Humphreys, R.M.\ 2003, \aj, 125, 1958
\bibitem[]{} McWilliam, A. \& Rich, R.M. 1994, ApJS, 91, 749
\bibitem[]{} Majewski, S.R.\ 1992, \apjs, 78, 87
 \bibitem[]{} Majewski, S., Skrutskie, M., Weinberg, M. \& Ostheimer, J.\ 2003, ApJ, 599, 1082
\bibitem[]{} Maller, A. \& Dekel, A.\ 2002, \mnras, 335, 487
\bibitem[]{} Maller, A., Katz, N., Keres, D. Dav\'e, R. \& Weinberg, D.\ 2006, \apj, 647, 763
\bibitem[]{} Matteucci, F. \& Brocato, E.\ 1990, \apj, 365, 539 
\bibitem[]{} Momany, Y. et al.\ 2006, A\&A, 451, 515 
 \bibitem[]{} Moore, B. et al.~1999, \apj, 524, L19
\bibitem[]{} Morrison, H., Flynn, C. \& Freeman, K.C.\ 1990, AJ, 100, 1191
\bibitem[]{} Mould, J. 2005, AJ, 129, 698
\bibitem[]{} Nordstr\"om, B. et al.\ 2004, A\&A, 418, 989
\bibitem[]{} Norris, J.E. \& Ryan, S.G.\ 1991, \apj, 380, 403
\bibitem[]{} Ortolani, S. et al.\ 1995, Nature, 377, 701

\bibitem[]{} Pagel, B.E.J. \& Patchett, B.E.\ 1975, \mnras, 172, 13
\bibitem[]{} Phleps, S., Meisenheimer, K., Fuchs, B. \& Wolf, C.\ 2000, A\&A, 356, 108
\bibitem[]{} Ratnatunga, K. \& Freeman, K. 1989, 339, 126 
\bibitem[]{} Reid, N. \& Majewski, S.R.\ 1993, \apj, 409, 635
\bibitem[]{} Rich, R.M.\ 1999, in `The Formation of Galactic Bulges', eds C.M.~Carollo, H.C.~Ferguson \&  R.F.G.~Wyse (CUP, Cambridge) p54
\bibitem[]{} Rix, H.-W. \& Zaritsky, D.\ 1995, ApJ,  447, 82
\bibitem[]{} Robertson, B. et al.\ 2006, \apj, 645, 986
\bibitem[]{} Rocha-Pinto, H.J. \& Maciel, W.J.\ 1996,  \mnras, 279, 447
\bibitem[]{} Rocha-Pinto, H.J., Scalo, J., Maciel, W.J. \& Flynn, C.\ 2000, A\&A, 358, 869
\bibitem[]{} Ryan, S. \& Norris, J.E. 1991, AJ, 101, 1865
\bibitem[]{} Sanchez, A.G. et al.\ 2006, \mnras, 366, 189
\bibitem[]{} Sandage, A.\ 1970, \apj, 162, 841
\bibitem[]{} Sandage, A. \& Fouts, G.\ 1987a, \aj, 93, 74
\bibitem[]{} Sandage, A. \& Fouts, G.\ 1987b, \aj, 93, 592
\bibitem[]{} Sandage, A. \& Katem, B.\ 1977, \apj, 215, 62
\bibitem[]{} Sargent, M.T. et al.\ 2006, \apj, in press (astro-ph/0609042)
\bibitem[]{} Schmidt, M.\ 1963, \apj, 137, 758
\bibitem[]{} Searle, L. \& Sargent, W.\ 1972, \apj, 173, 25
\bibitem[]{} Siegel, M.H., Majewski, S.R., Reid, N. \& Thompson, I.B.\ 2002, \apj, 578, 151
\bibitem[]{} Spergel, D.\ et al.~(WMAP team) 2006, ApJ in press (astro-ph/0603449) 
\bibitem[]{} Steinmetz, M.~et al.\ 2006, \aj, 132, 1645
\bibitem[]{} Stephens, A. \& Boesgaard, A.\ 2002, \aj, 123, 1647
\bibitem[]{} Tinsley, B.\ 1975, \apj, 197, 159
\bibitem[]{} Tinsley, B.\ 1976, \apj, 208, 797
\bibitem[]{} Tinsley, B.\ 1977, \apj, 211, 621
\bibitem[]{} Tinsley, B.\ 1979, \apj, 229, 1046
\bibitem[]{} Tolstoy, E. et al.\ 2003, AJ, 125, 707
\bibitem[]{} Unavane, M., Wyse, R.F.G. \& Gilmore, G.\ 1996, MNRAS, 278, 727
\bibitem[]{} van den Bergh, S.\ 1962, \aj, 67, 486
\bibitem[]{} van den Bergh, S.\ 1980, in `Scientific Research with the Space Telescope', IAU Colloquium 54, eds M.S.~Longair \& J.W.~Warner (NASA, CP-2111) p151
\bibitem[]{} van Loon, J.~et al. 2003, MNRAS, 338, 857
\bibitem[]{} Venn, K., Irwin, M., Shetrone, M., Tout, C., Hill, V. \& Tolstoy, E.\ 2004, AJ, 128, 1177
\bibitem[]{} Vivas, A.K. \& Zinn, R.\ 2006, \aj, 132, 714

\bibitem[]{} Weil, M., Eke, V. \& Efstathiou, G.\ 1998, \mnras, 300, 773
\bibitem[]{} Wilkinson, M. et al.\ 2006, in proc.~XX1st IAP Colloquium, EAS Publications Series, Vol.~20, p105 (astro-ph/0602186)

\bibitem[]{} Wyse, R.F.G. 1999, in `The Formation of Galactic Bulges',
eds C.M.~Carollo,
H.C.~Ferguson \& R.F.G.~Wyse (CUP, Cambridge) p195

\bibitem[]{} Wyse, R.F.G.\ 2001, in `Galactic Disks and Disk Galaxies' ASP Conference Series, Vol. 230,  
eds.\ J.G.~Funes, S.J. \& E.M.~ Corsini (San Francisco: ASP), p71
\bibitem[]{} Wyse, R.F.G. \& Gilmore, G.\ 1986, AJ, 91, 855
\bibitem[]{} Wyse, R.F.G. \& Gilmore, G. 1988, AJ, 95, 1404
\bibitem[]{} Wyse, R.F.G. \& Gilmore, G.\ 1989, ComAp, 13, 135
\bibitem[]{} Wyse, R.F.G. \& Gilmore, G. 1992, AJ, 104, 144
\bibitem[]{} Wyse, R.F.G. \& Gilmore, G.\ 1995, AJ, 110, 2771
\bibitem[]{} Wyse, R.F.G., Gilmore, G., Houdashelt, M., Feltzing, S.,
Hebb, L., Gallagher, J. \& Smecker-Hane, T. 2002, New Astr, 7, 395
\bibitem[]{} Yanny, B.~et al.\ 2003, ApJ, 588, 824
\bibitem[]{} Yoachim, P. \& Dalcanton, J. 2005, ApJ, 624, 701
\bibitem[]{} Yoshii, Y.\ 1982, PASJ, 34, 365
\bibitem[]{} Zoccali, M.~et al. 2000, ApJ, 530, 418
\bibitem[]{} Zoccali, M. et al.\ 2006, A\&A, in press (astro-ph/0609052)

\end{thebibliography}

\end{document}